\documentclass[a4paper,fleqn]{cas-sc}

\usepackage[numbers]{natbib}
\usepackage{subcaption}
\usepackage{amsmath,amssymb,amsfonts}
\usepackage{algorithmic}
\usepackage{graphicx}
\usepackage{textcomp}
\usepackage{xcolor}
\def\BibTeX{{\rm B\kern-.05em{\sc i\kern-.025em b}\kern-.08em
    T\kern-.1667em\lower.7ex\hbox{E}\kern-.125emX}}
    
\usepackage{afterpage} 
\usepackage[ruled,vlined]{algorithm2e} 
\usepackage{booktabs} 

\def\tsc#1{\csdef{#1}{\textsc{\lowercase{#1}}\xspace}}
\tsc{WGM}
\tsc{QE}
\tsc{EP}
\tsc{PMS}
\tsc{BEC}
\tsc{DE}

\begin{document}
\let\WriteBookmarks\relax
\def\floatpagepagefraction{1}
\def\textpagefraction{.001}

\shorttitle{Guaranteeing Anonymity in Attribute-Based Authorization}

\shortauthors{Lanus, Colbourn, and Ahn}

\title [mode = title]{Guaranteeing Anonymity in Attribute-Based Authorization}                      
\tnotemark[1]

\tnotetext[1]{Research of EL was supported by a National Physical Science Consortium (NPSC) Fellowship. Research of CJC was supported in part by the National Science Foundation under Grant No. 1421058 and Grant No. 1813729.}

%
\author[1]{Erin Lanus}

\cormark[1]


\ead{lanus@vt.edu}



\affiliation[1]{organization={Virginia Tech National Security Institute, Virginia Tech},
    city={Arlington},
    state={VA},
    country={USA}}

\author[2]{Charles J. Colbourn}

\author[2]{Gail-Joon Ahn}


\affiliation[2]{organization={School of Computing and Augmented Intelligence, Arizona State University},
    city={Tempe},
    state={AZ},
    country={USA}}

\cortext[cor1]{Corresponding author}



\begin{abstract}
Attribute-based methods, such as attribute-based access control and attribute-based encryption, make decisions based on attributes possessed by a subject rather than the subject's identity. While this allows for anonymous authorization -- determining that a subject is authorized without knowing the identity of the subject -- it does not guarantee anonymity. If a policy can be composed such that few subjects possess attributes satisfying the policy, then when the policy is used for access control, in addition to making a grant or deny decision, the system can also guess with high probability the identity of the subject making the request. Other approaches to achieving anonymity in attribute-based authorization do not address this attribute distribution problem. Suppose polices contain conjunctions of at most $t$ attributes and the system must not be able to guess with probability greater than $\frac{1}{r}$ the identity of a subject using a policy for authorization. We say the anonymity guarantee is $r$ for maximum credential size $t$. An anonymizing array is a combinatorial array proposed as an abstraction to address the underlying attribute distribution problem by ensuring that any assignment of values to $t$ attributes appearing in the array appears at least $r$ times. Anonymizing arrays are related to covering arrays with higher coverage, but have an additional desired property, homogeneity, due to their application domain. In this work, we discuss the application of anonymizing arrays to guarantee anonymous authorization in attribute-based methods. Additionally, we develop metrics, local and global homogeneity, to compare anonymizing arrays with the same parameters.
\end{abstract}




\maketitle

\section{Introduction}

\subsection{Attribute-Based Systems}
Attribute-based access control (ABAC) is a logical access control model wherein access control decisions are made on the basis of attributes. The National Institute of Standards and Technology (NIST) defines ABAC as ``an access control method where subject requests to perform operations on objects are granted or denied based on assigned attributes of the subject, assigned attributes of the object, environment conditions, and a set of policies that are specified in terms of those attributes and conditions" \cite{hu2013guide}. Attributes are characteristics of a subject expressed as name-value pairs and may be based on the subject's real-world identity (e.g., $name=Martha$, $age=34$) or in-system characteristics (e.g., $role=administrator$). ABAC can be configured to support policies available under other models of access control such as discretionary access control, mandatory access control, and role-based access control (RBAC)\cite{hu2013guide}. One feature that makes ABAC attractive is that it can be used as global access control on a heterogeneous system where subsystems are each employing a different access control model. Similarly, hybrid models have been proposed, such as adding attributes to RBAC \cite{kuhn2010adding}. 

Unlike other access control models, ABAC is inherently identity-less. Determining whether to issue a grant or deny response to an access request for a resource is characteristically done on the basis of the attributes the subject possesses and presents via some credential; it does not require knowledge of the identity of the subject. This may greatly simplify privilege management. Making decisions on the basis of attributes rather than identities removes the need to update access control lists or capability lists to add or remove privileges related to individual subjects. Policies can be written dynamically to achieve fine-grained access control without creating roles for each possible subset of users in advance, resulting in a ``role explosion'' \cite{kuhn2010adding}. ABAC is desirable in systems with a large number or frequently changing set of users.

Attribute-Based Encryption (ABE) is another general attribute-based method. ABE is first introduced as Fuzzy Identity-Based Encryption (FIBE) where identities are sets of attributes. In FIBE, if an identity matches on $d$ or more of the attributes used to encrypt a ciphertext, the private key affiliated with the identity can be used to decrypt \cite{sahai2005fuzzy}. The access structure employed by FIBE is restricted to a threshold gate the size of which is fixed at setup time. In Key-Policy ABE, private keys are access control policies and ciphertexts are encrypted over attributes \cite{goyal2006attribute}. A private key that contains a policy that is satisfied by the attributes of a ciphertext can be used to decrypt the ciphertext. Conversely, Ciphertext-Policy ABE (CP-ABE) encrypts ciphertexts with access control policies \cite{bethencourt2007ciphertext}. A private key that contains attributes that satisfy the policy can be used to decrypt.

CP-ABE has been proposed to mediate authenticated key exchange \cite{portnoi2016location}. As a simplified example, suppose the service encrypts a session key in a policy and broadcasts the message. A subject that has a private key containing attributes that satisfy the policy is able to decrypt the message and obtain the key. The subject begins communicating with the service by encrypting messages in the session key. The service now knows that the subject communicating via the session key is authorized based on possession of the required attributes to obtain the key. 

\subsection{Privacy and Anonymity in Access Control} 

Privacy-preservation is an important topic in access control due to the integration of information systems into nearly all aspects of our lives, the amount of personal data that is collected and stored, and laws governing how information is to be used and secured. Different scenarios emphasize preserving privacy of essentially three different entities and lead to different solutions. The most common privacy concern is protecting sensitive data stored on an access-controlled system, such as an individual’s personally identifiable information. Privacy-aware role-based access control (P-RBAC) is a family of models developed to extend RBAC permission assignments to support privacy policies and protect sensitive data in terms of purposes, conditions, and obligations \cite{ni2010privacy}. A privacy-aware access control module using user declarations and credentials as certificates to control secondary use of personal information is described in \cite{ardagna2008privacy}. Second, privacy of the system or the encryptor leads to solutions that seek to hide the access control policy \cite{nishide2008attribute}. The third case, identity privacy of the subject requesting access or subject anonymity, is the focus of this work. Consider a business utilizing cloud computing. Even when data privacy against the service is achieved by operating on encrypted data, the service may collect information from subject access histories, infer sensitive information about the data content, and sell the leaked information to a competitor. Ideally, inference channels are prevented by achieving subject anonymity even when attributes are public and known to the system.

A feature of attribute-based systems is that they can be used to achieve anonymous access control, deciding the authorization status of subjects without requiring knowledge of the identity of the subject. This feature is natural to systems implementing pure ABAC as ``in many models, access requests are not conducted directly by the user but indirectly through a session that may contain a subset of the user's attributes'' \cite{servos2017current}. In some instances, the subject uses an application that allows the subject to choose which attributes to present to the system. In others, a credential such as a card preloaded with a subset of attributes represents the subject to the system. Unless the subject's identity is used as an attribute or the system requires the subject to send subject-unique information, anonymous access control is a possible byproduct of ABAC. 

While desirable to subjects, anonymity is not always considered to be a feature due to the potential conflict between anonymity and auditability. Whether these two goals conflict is dependent upon the definition of auditability for the system. One definition of auditability is ``the ability to easily determine the set of users who have access to a given resource or the set of resources a given user may have access to (sometimes referred to as a `before the fact audit'),'' \cite{servos2017current}. In our proposal and to model a system with the most  adversarial power to break subject identity privacy, we assume that all attributes of all subjects are registered with the system. The set of subjects possessing attributes that satisfy a given policy can be computed by the system to determine if the policy is appropriate for a given resource or if it needs to be improved. Additionally, the system can log access requests along with the credential presented and the access control decision to check if unauthorized attempts were made. Auditability in this sense is achievable concurrently with anonymity.
 However, if the goal of auditability is to be able to identify the subject who completed an operation, anonymity against the system and auditability are necessarily at odds. The purpose of this work is to demonstrate that the degree of anonymity achieved for subjects when attribute-based methods are used for authorization is directly related to the distribution of attributes over subjects. A dual outcome is that we also show when a system employing attribute-based methods cannot guarantee to identify subjects gaining authorization to the system.

\subsection{Related Work}
Whether anonymity is achieved and even the kind of anonymity intended is typically dependent upon the ABAC implementation. One approach is to provide certificates to prove possession of attributes. Still, transactions involving the same certificates may be linked, the certificates may require the subject's public key certificate, or the approach may require sending all certificates for potentially relevant attributes \cite{backes2005anonymous}. The solution in \cite{backes2005anonymous} employs cryptographic zero knowledge proofs to allow subjects to prove possession of just the requisite attributes without revealing additional information. The assumption in \cite{kolter2007privacy} is that a subject may not wish to disclose private attributes, such as a credit card number, to a particular system, and they propose an architecture that requires a trusted third party sit between the subject and an ABAC-protected system. In addition to effectively moving the requirement of trust from the system to a third party, a challenge with this approach informed by the work in $k$-anonymity is that individual or even sets of attributes may not be considered sensitive until aggregated. Another approach splits a subject's attributes among multiple authorities, none of which know all of the subject's attributes and thus relies on anonymity through obscuring the attributes \cite{jung2014control}.

ABE provides another mechanism for anonymous authorization. LOCATHE, a system employing CP-ABE, includes a mode in which the subject does not provide its identity to the system \cite{portnoi2016location}. Instead, the service knows the subject is authorized based on possession of attributes. The authors of LOCATHE make the stronger claim that the service cannot uniquely identify the subject. While attribute-based methods typically do not require knowledge of the identity of the subject in order to make an authorization decision, this is not a guarantee that a system cannot determine the subject's identity. For example, in the LOCATHE system, the subject must register with the service running LOCATHE to receive a key, and thus the service knows all of the users in the system as well as their attributes. If the system can compose a policy that only one subject possesses the attributes to satisfy and encrypts the session key in this policy, the service knows with certainty the identity of the subject communicating with it. ``Anonymous ABE'' uses hidden credentials which can be used to retrieve a session key anonymously. However, anonymity on the part of the key receiver is based on ``plausible deniability'' as anyone can request the encrypted message with the key, not just the intended receiver \cite{kapadia2007attribute}. Someone with the correct credentials must decrypt the message to gain the session key to obtain authorization, so plausible deniability fails with use of the session key.

$k$-anonymous attribute-based access control utilizes a cryptographic private matching protocol for a service to convince a subject that if the subject submits a set of attributes, the service has already received $k$ previous requests with identical attributes \cite{squicciarini2007kanonymous}. Here, credentials are sets of attributes signed by an authority. The service proves that it possesses valid previous requests to the degree of anonymity, $k$, specified by the subject. The subject proves that it possesses valid credentials without leaking additional information. Like ABE, this is a mechanism for communicating possession of attributes between subject and system, but it requires that the system has received $k$ previous requests. Thus, the underlying attribute distribution problem exists also for and is not solved by $k$-anonymous attribute-based access control.

\subsection{Contribution}
The primary contribution of this work is to demonstrate that guarantees of anonymity in attribute-based authorization depend on certain properties of attribute distribution. That is, when these properties of attributes hold, a system granting access to authorized subjects and denying it to unauthorized subjects cannot guess the identities of subjects by the credentials presented. 
Without loss of generality, consider policies to be disjunctions of conjunctions of attribute values. The most restrictive policy is a single conjunction of many attribute values. Let the \emph{maximum credential size} used by a system contain $t$ attributes, so polices contain conjunctions of at most $t$ attributes. Suppose the system must not be able to guess with probability greater than $\frac{1}{r}$ the identity of a subject presenting a credential to satisfy a policy for an access request. Then the \emph{anonymity guarantee} is $r$. An \emph{anonymizing array} is a combinatorial array proposed as an abstraction to address the underlying attribute distribution problem by ensuring that any assignment of values to $t$ attributes appearing in the array appears at least $r$ times.

When the set of subject attributes registered to a system is fixed, an anonymizing array determines the maximum credential size that can be used while achieving the anonymity guarantee $r$ or, equivalently, the guarantee achievable for a given size $t$. When the set of registered attributes can be appended, anonymizing arrays provide a mechanism to provide higher anonymity guarantees. A key benefit of anonymizing arrays is that they are implementation independent and can be used to enhance subject anonymity of existing systems employing ABAC or using ABE as an access control mechanism. As opposed to $k$-anonymous ABAC, our model assumes that all subjects and attributes are known to the system and the anonymity guarantee $r$ is available immediately. 

The rest of the paper is organized as follows. Definitions and how to compute the anonymity guarantee are in \S~\ref{section:AAs}. An additional desired property of anonymizing arrays and metrics for comparing two anonymizing arrays with the same parameters on this property are presented in \S~\ref{section:homogeneity}. Discussion of how this problem differs from the body of work on $\kappa$-anonymity in statistical databases and future directions of practical interest such as implementing least privilege with anonymity are explored in \S~\ref{section:discussion}. Conclusions are given in \S~\ref{section:conclusion}.

\section{Anonymizing Arrays} \label{section:AAs}
\subsection{Definition}

Consider an array with $N$ rows and $k$ columns and each column $i$ has entries from a set of $v_i$ symbols for $1 \leq i \leq k$. Such an array is \emph{$(r,t)$-anonymous} if, for every way to pick $t$ columns, every $t$-tuple of symbols appearing in those columns appears at least $r$ times. Given an $N \times k$ array $\mathbf{A}$ and an $N' \times k$ array $\mathbf{A'}$, if $\mathbf{A} \subseteq \mathbf{A'}$ and $\mathbf{A'}$ is $(r,t)$-anonymous, $\mathbf{A'}$ is \emph{$(r,t)$-anonymizing} with respect to $\mathbf{A}$. To write the parameters of an anonymizing array, we use the notation \textsf{AA}$(N; r, t, k, (v_1,\ldots, v_k))$. When $j$ columns share the same number of symbols $v_i$, exponential notation $v_{i}^j$ is used for simplicity. The order of the columns and order of the rows is not important for the properties to hold. 

Once an anonymizing array is discussed in the context of an attribute-based system, we refer to a row as an \emph{access profile} and columns as \emph{attributes} with the symbols in the array as attribute values. Then, an access profile is an assignment of values to all attributes. A \emph{credential} is a tuple of up to $t$ attribute-value pairs. Again, $t$ is the maximum credential size with $1 \leq t \leq k$, and $r$ is the anonymity guarantee. The anonymity guarantee is trivial when $r = 1$ as every credential that appears in an array appears at least once by definition; interesting cases require $r > 1$. Attributes with only one value are trivial as every row contains that attribute, so useful attributes require at least two values. A consequence of this approach is that the numbers of attribute values must be finite. If new values must be added, the prior anonymity guarantee may no longer hold. The array parameters are summarized in Table~\ref{table:parameters}.

\begin{table}[b]
    \centering
    \caption{Anonymizing Array Parameters}
    \begin{tabular}{c|l}
        \toprule
        Parameter & Meaning\\
        \midrule
        $N$ & number of access profiles (rows) \\
        $k$ & number of attributes (columns)\\
        $v_i$ & number of values for attribute $i$ (symbols)\\
        $r$ & anonymity guarantee (repetitions of $t$-set) \\
        $t$ & size of credential (number of columns)\\
        \bottomrule
    \end{tabular}
    \label{table:parameters}
\end{table}

Access profiles may correspond to a subject or may exist in the system as \emph{padding}, ``dummy'' rows created for the purpose of reaching the anonymity guarantee as subjects interact with the system by presenting sets of attributes. Because padding rows are added to reach a degree of anonymity not provided by the subject access profiles, care should be taken to construct rows that are not obviously padding. Access profiles need not be unique. 

A user credential is defined in \cite{servos2017current} as ``sets of typed attributes relating to the same topic or structure, e.g., an employee  credential may contain an age, address,  and salary  attribute,'' and users may select the subset of attributes activated in a session. The term credential typically has a stronger connotation than just a random set of attributes, including some mechanism by which the subject proves possession of those attributes. This may differ across implementations, from providing a physical card loaded with exactly that set of attributes to an application that allows the subject to decide which attributes to provide and which to withhold. For the abstraction to be implementation agnostic, the term credential is used generically to mean the tuple of attribute and attribute value pairs that are being provided for a given request. Authorization decisions are made by the system based on the credential presented rather than the access profile. 

Given a set of attributes, each access profile has exactly one credential that corresponds to an assignment of values to those attributes. In context, an array is $(r,t)$-anonymous if, for all credentials of $t$ or fewer attributes, there are at least $r$ identical credentials distributed among the $N$ access profiles. This provides the anonymity guarantee that, if a subject presents a credential of $t$ or fewer attributes in an access control scenario, there are at least $r-1$ other access profiles corresponding to the credential. Thus, the system can identify the subject with not greater than $\frac{1}{r}$ probability.

\subsection{Constraints}
\emph{Hard constraints} are credentials that must not appear, while \emph{soft constraints} are credentials that are not forbidden but need not appear. Avoiding violation of constraints is important when adding padding rows to an anonymizing array in order to meet the anonymity guarantee. Some attribute assignments may be impossible. For example, consider employee roles that only exist for a certain facility. One attribute might represent where the employee is stationed, while another is the employee's role. An employee not at facility F may not be assigned role R. This assignment should not exist in the access profiles provided, and appearance of this combination immediately identifies the containing access profile as padding. This is a hard constraint. Other assignments may not be impossible and yet do not appear in the provided set of access profiles. They can be included in padding rows, but if they are included, they must appear $r$ times. These are soft constraints. Every credential not specified in a hard or soft constraint must appear $r$ times. 

The case where no feasible solution exists can arise when hard constraints lead to \emph{implicit hard constraints}. For example, consider building an anonymizing array \textsf{AA}$(N;2, 2, 3, (2^3))$ with attribute values as $\{0,1\}$. If the only constrained credentials are hard constraints $\{(a_1,0),(a_2,0)\}$ and $\{(a_1,0),(a_2,1)\}$, then any credential containing $(a_1$,0) is implicitly constrained. There is no way to assign a value to attribute $a_2$ if attribute $a_1$ has value $0$ without violating the constraints. But any credential not given as a hard or soft constraint must appear twice, so there is no solution. Table~\ref{table:constraints} enumerates the unconstrained credentials as those that do not appear in a constraint and shows in bold the unconstrained credentials that are implicitly constrained by the hard constraints for this example. If the implicit constraints are defined as constraints themselves, a solution that does not violate the hard constraints exists and is given in Figure~\ref{constraintAA}. If the hard constraints are instead soft constraints, then a solution also exists, as these credentials can appear as needed to ensure unconstrained credentials appear, provided they appear twice. (See Figure~\ref{fig:full8} in \S \ref{section:homogeneity} for an example).

\begin{table}
\begin{center}
\caption{Implicit hard constraints}
\label{table:constraints}
\begin{tabular}{c c}
\toprule
Hard Constraints & Unconstrained Credentials\\
\midrule
\{$(a_1$,0),($a_2$,0)\} 	& \{$(a_1$,1),($a_2$,0)\} \\
\{$(a_1$,0),($a_2$,1)\}	& \{$(a_1$,1),($a_2$,1)\} \\
& \boldmath${\{(a_1,0),(a_3,0)\}}$\\
& \boldmath${\{(a_1,0),(a_3,1)\}}$\\
& \{$(a_1$,1),($a_3$,0)\} \\
& \{$(a_1$,1),($a_3$,1)\} \\
& \{($a_2$,0),($a_3$,0)\} \\
& \{($a_2$,0),($a_3$,1)\} \\
& \{($a_2$,1),($a_3$,0)\} \\
& \{($a_2$,1),($a_3$,1)\} \\
\bottomrule
\end{tabular}
\end{center}
\end{table}

\begin{figure}
\begin{center}
\begin{tabular}{l | c c c}
\toprule
& $a_1$ & $a_2$ & $a_3$\\
\midrule
1 & 1 & 0 & 0\\
2 & 1 & 0 & 0\\
3 & 1 & 0 & 1\\
4 & 1 & 0 & 1\\
5 & 1 & 1 & 0\\
6 & 1 & 1 & 0\\
7 & 1 & 1 & 1\\
8 & 1 & 1 & 1\\
\bottomrule
\end{tabular}
\caption{\textsf{AA}$(8;2,2,3,(2^3))$ with hard constraints}
\label{constraintAA}
\end{center}
\end{figure}


\subsection{Relationship to Covering Arrays}
A \emph{covering array} denoted \textsf{CA}$(N;t,k,v)$ is an array with $N$ rows and $k$ columns on $v$ symbols such that for every set of $t$ columns each of the $v^t$ combinations of symbols appears in at least one row of the array. These combinations are called \emph{interactions} and are said to be \emph{covered}. Covering arrays are employed to construct test suites in combinatorial interaction testing to ensure that the suite includes a test for any fault caused by an interaction of up to $t$ components in the system under test. Anonymizing arrays are similar to covering arrays with constraints and requiring a higher level of coverage. The application domain of anonymizing arrays, however, leads to differences in how constraints are determined, and they have an additional desired property, homogeneity, defined in \S~\ref{section:homogeneity}. A detailed discussion of the relationship between covering and anonymizing arrays as well as the  adaptation of some covering array construction algorithms for anonymizing arrays is in \cite{lanus2020algorithms}.

\subsection{Anonymizing Array Example}

Consider a system at a university that has the following attributes and values:
\begin{enumerate}
\item $Role$=\{$faculty$, $graduate$, $undergraduate$\}
\item $Job$=\{$instructor$, $grader$\}
\item $Department$=\{$CS$, $EE$\}
\item $Semester$=\{$Spring$, $Fall$\}
\end{enumerate}
The access profiles provided to the system are in array $\mathbf{A}$ (Figure~\ref{figure:A}). Suppose the system requires the hard constraints ($faculty$, $grader$) and ($undergraduate$, $instructor$). $\mathbf{A}$ also has the soft constraint ($graduate$, $grader$). Every unconstrained pair of attributes appears at least once, so $\mathbf{A}$ is $(1,2)$-anonymous. It is not $(2,2)$-anonymous as only one access profile has the credential ($CS$, $grader$). Array $\mathbf{B}$ is $(2,2)$-anonymizing for $\mathbf{A}$. The first six access profiles of $\mathbf{B}$ are $\mathbf{A}$ and then six padding rows have been added. It is $(2,2)$-anonymous because every pair of attributes appears at least twice. Twelve rows are required, as the largest number of unconstrained credentials for a pair of attributes is six and each must appear twice. The soft constraint appears in $\mathbf{B}$ twice, but another anonymizing array for $\mathbf{A}$, $\mathbf{B'}$, exists with the credential ($graduate$, $instructor$) in rows 9 and 10 instead. $\mathbf{B}$ is not $(3,2)$-anonymous, as the credential $(faculty, CS)$ does not appear at least three times. It is not $(2,3)$-anonymous, as the credential $(graduate, grader, Fall)$ appears only once. If $graduate \land grader \land Fall$ is a policy used for an authorization decision, the access profile in row 9 is uniquely identified. 

\begin{figure}[t]
\begin{center}
\begin{tabular}{r | l l l l}
\toprule
	& 	Role	 & 	Job	& 	Department	& 	Semester	\\
	\midrule
1	& 	faculty	& 	instructor	& 	CS	& 	Spring	\\
2	& 	faculty	& 	instructor	& 	EE	& 	Fall	\\
3	& 	graduate	& 	instructor	& 	CS	& 	Spring	\\
4	& 	graduate	& 	instructor	& 	EE	& 	Fall	\\
5	& 	undergraduate	& 	grader	& 	CS	& 	Fall	\\
6	& 	undergraduate	& 	grader	& 	EE	& 	Spring	\\
\bottomrule
\end{tabular}
\caption{Array A, an $\textsf{AA}(6;1,2,4,(3,2^3))$}
\label{figure:A}
\end{center}
\end{figure} 

\begin{figure}[t]
\begin{center}
\begin{tabular}{r | l l l l}
\toprule
	& 	Role	& 	Job	& 	Department	& 	Semester	\\
	\midrule
1	& 	faculty	& 	instructor	& 	CS	& 	Spring	\\
2	& 	faculty	& 	instructor	& 	EE	& 	Fall	\\
3	& 	graduate	& 	instructor	& 	CS	& 	Spring	\\
4	& 	graduate	& 	instructor	& 	EE	& 	Fall	\\
5	& 	undergraduate	& 	grader	& 	CS	& 	Fall	\\
6	& 	undergraduate	& 	grader	& 	EE	& 	Spring	\\
7	& 	faculty	& 	instructor	& 	CS	& 	Fall	\\
8	& 	faculty	& 	instructor	& 	EE	& 	Spring	\\
9	& 	graduate	& 	grader	& 	CS	& 	Fall	\\
10	& 	graduate	& 	grader	& 	EE	& 	Spring	\\
11	& 	undergraduate	& 	grader	& 	CS	& 	Fall	\\
12	& 	undergraduate	& 	grader	& 	EE	& 	Spring	\\
\bottomrule
\end{tabular}
\caption{Array B, an $\textsf{AA}(12;2,2,4,(3,2^3))$}
\label{figure:B}
\end{center}
\end{figure} 
\subsection{Computing the Anonymity Guarantee}

As the credential size increases, the number of repetitions of the credential in an array must either stay the same or decrease, so an inverse relationship exists between $t$ and $r$.  Consider an array $\mathbf{A}$ for a given $t$ so the maximum anonymity guarantee available is $r$. By definition, $\mathbf{A}$ is $(r,t)$-anonymous and is not $(r+1,t)$-anonymous. Let $c$ be a credential over $t$ attributes with exactly $r$ appearances in $\mathbf{A}$. Look at the set of rows in which $c$ appears. Choose any credential  $c'$ of size $t' > t$ such that $c'$ contains $c$. The rows in which $c'$ appears must be a subset of the rows in which $c$ appeared. Then for every $t'$ with $t < t' \leq k$ for which $\mathbf{A}$ is $(r',t')$-anonymous, it must be the case that $r' \leq r$.  Additionally, pick any credential, $c$, of size $t$. It appears in at least $r$ rows. Then any $t''$-subset of $c$ appears in at least these rows. Consequently, an array that is $(r,t)$-anonymous is $(r,t'')$-anonymous for every $t'' < t$. The practical consequence is that $r$ and $t$ are tunable given the privacy needs of users in a system. 

Given an anonymizing array $\mathbf{A}$ and a maximum credential size $t$, Algorithm~\ref{alg:guarantee} computes the maximum $r$ for which $\mathbf{A}$ is $(r,t)$-anonymous. It also serves to check an anonymizing array. If a hard constraint violation is found, it returns 0. Recall that soft constraints may appear zero or at least $r$ times. To compute all pairs $r$ and $t$ for $\mathbf{A}$,  Algorithm~\ref{alg:guarantee} can be executed for increasing values of  $t$ until it returns $r=1$.

 Algorithm~\ref{alg:guarantee} systematically checks the array one $t$-set at a time and scans each row within an $N \times t$ subarray only once; thus, it never looks at the same $t$-set of cells more than once. If a system allows any of the $\binom{k}{t}$ $t$-sets of attributes to be used in a policy, the array must anonymize for each and thus all must be checked. If some $t$-sets of attributes are not allowed in a policy, the for loop can be restricted to only consider the allowable $t$-sets.

\begin{algorithm}
\DontPrintSemicolon
\SetKwInOut{Input}{input}\SetKwInOut{Output}{output}
\Input{ $\mathbf{A}$ an $N \times k$ array, $t$}
\Output{$r$}
\Begin
{
	\For{each of the $\binom{k}{t}$ sets of columns of $\mathbf{A}$}
	{
		Scan the $N \times t$ subarray of $\mathbf{A}$ and store the count of each credential\\
		
		\If{a hard constraint has $count > 0$}
		{
			Return 0
		}
		\Else
		{
			Set $r$ to be the smallest non-zero count so far\\
		}
	}
	Return $r$
}
\caption{Compute Anonymity Guarantee \label{alg:guarantee}}
\end{algorithm}
\section{Homogeneity Metrics}\label{section:homogeneity}
 
\subsection{Designing Metrics}
In addition to requiring that all credentials appear either zero or at least $r$ times, we prefer anonymizing arrays that do not contain groups of highly similar access profiles. If an access profile shares all of its ${k \choose t}$ credentials exclusively with a group of $r-1$ other access profiles, when any of these credentials is used, the system can identify someone in the group as the active subject. Even if no one subject can be identified with greater than $\frac{1}{r}$ probability, this is undesirable as it allows access requests using these credentials to be linked and behavior of the group to be tracked. An anonymizing array in which access profiles share different credentials with different groups is preferable to another anonymizing array with the same anonymity guarantee in which access profiles share credentials with the same groups frequently. Lastly, the size of the groups matters as well. Tracking the behavior of a group of $2r$ access profiles that share all credentials is less targeted to one access profile than tracking the behavior of a group of $r$ access profiles. An ideal solution captures the interplay of these two concerns in a single metric.

Consider a multi-hypergraph representation of an anonymizing array. There are $N$ vertices, one for each of the access profiles. Each vertex has degree ${k \choose t}$, and a hyperedge represents the credential possessed by the access profile for this $t$ set of attributes connecting all of the access profiles that share this credential. When all attributes have $v$ values, there are up to ${k \choose t}v^t$ edges. (When attributes have different numbers of values, replace $v^t$ with $\prod_{j=1}^{t}v_j$ for each of the ${k \choose t}$ sets of attributes. We use constant $v$ for simplicity in counting throughout this section, but all of the results extend naturally.) It is a multigraph as the same vertices can be connected by multiple hyperedges if the same group shares more than one credential. The desired metric, then, measures the number of edges that connect a vertex to the same set of vertices and the sizes of those sets. 

A clustering coefficient can be described as the measure of neighbors of a vertex that know each other; if vertex $v$ has $m$ neighbors, the clustering coefficient is the fraction of the $m(m-1)/2$ possible edges that actually exist between those vertices \cite{watts1998collective}. The clustering coefficient describes the tendency of vertices to form cliques. In a scenario where all $N$ access profiles share a universal credential, the clustering coefficient masks any other behavior exhibited in the rest of the graph. Additionally, work to extend this concept to hypergraphs ignores multi-edges, considering them a simple edge \cite{estrada2005complex}. The desired metric must measure how many times a vertex is connected to a group of vertices, not just how many vertices it is connected to, so the multi-edges cannot be ignored. Structural similarity has the same issue. It measures how many neighbors two vertices have in common, but not how often they have similar neighbors. 

Applying the Jaccard index, or intersection over union, to neighbors of a vertex can be misleading \cite{jaccard1908nouvelles}. Two vertices may share edges with many of the same vertices and so have very similar neighborhoods, and yet may appear infrequently in edges together. See Figure~\ref{lowgraph} for an example. Access profiles 1 and 2 share an edge, and yet have a Jaccard index of 0 if a vertex is not included as its own neighbor; if it is, the index is .33. Access profiles 1 and 4 do not share an edge, and yet have a Jaccard index of .5 if a vertex is not included as its own neighbor and .33 if it is.

We can also think of an anonymizing array as a set of ${k \choose t}$ hypergraphs each having copies of the same $N$ vertices and each graph having up to $v^t$ edges. To ask how similar the neighborhood of some vertex $u$ is across all of the graphs, a similarity score for neighborhoods is required. This might require counting the number of vertices in common and making ${{k \choose t} \choose 2}$ comparisons of the neighborhoods between graphs for each of the $N$ vertices. Defining the similarity score still presents difficulty, as the sets $\{u_i, u_j\} \bigcap \{u_i, u_j\}$ and $\{u_i, u_j, u_l, u_m\} \bigcap \{u_i, u_j, u_p, u_q, u_r, u_s\}$ have the same score, 2, under similarity by intersection. For our purposes, the first two sets are much more similar than the last two as every vertex that appears in one neighborhood appears in the other. Taking the Jaccard index here would distinguish these cases, but produces a score that ignores the size of the neighborhoods. That is, the Jaccard index of two sets containing the same two elements is 1, as is the Jaccard index for two sets containing the same 100 elements. As mentioned, appearing together in large groups is preferred to appearing together in small groups; taking intersection over union will not distinguish these cases.  No well-known graph measure accurately describes the metric required here.

\emph{Diversity} is a metric for covering arrays \cite{nie2015combinatorial} defined as $$\frac{\text{number of distinct $t$-way interactions covered}}{\text{number of opportunities to cover $t$-way interactions}}.$$ To be a covering array without constraints, the number of $t$-way interactions covered must be $\binom{k}{t}$. Diversity measures how efficiently interactions are covered. An array that covers all $t$-way interactions in fewer rows has a higher diversity score. The number of times a particular $t$-way interaction is covered is ideally close to one. 

For anonymizing arrays, the number of times a credential appears is ideally close to $r$. If some credentials appear many more times than $r$, it is possible that the anonymity guarantee can be increased if some of the repetitions of those credentials were replaced with less frequently appearing credentials. Superficially, it seems that diversity can be extended to anonymizing arrays as $$\frac{\text{number of distinct $t$-sized credentials covered at least $r$ times}}{\text{number of opportunities to cover $t$-sized credentials}}.$$ Dependence between rows does not matter for covering arrays as these are tests in a test suite that will be executed independently, but dependence between access profiles does matter for anonymizing arrays. Diversity does not capture the preference for two or more access profiles to not contain many of the same credentials.

Define the \emph{neighborhood} of a credential to be the access profiles that share that credential; this is equivalent to the set of vertices connected by a hyperedge, maintaining multi-edges as unique edges. Computing the neighborhood size for each credential measures how inclusive or exclusive the credential is, but does not say anything about which access profiles appear together in neighborhoods. Computing the total number of neighbors for an access profile, all neighbors of the vertex, says something about how many other access profiles it shares a credential with, but does not detect when a vertex has many small neighborhoods of the same neighbors and one large neighborhood shared with everyone versus many diverse neighborhoods. Computing the number of times a pair of access profiles appear together is also useful, but again ignores the effect of large versus small neighborhoods. To reiterate, a measure that includes the number of times access profiles are neighbors and the size of the neighborhoods is needed, and we propose a solution in the next section.

\subsection{Homogeneity Definitions}

The \emph{weight} of a pair of access profiles on a credential is zero if the access profiles do not share the credential and inversely proportional to the size of the neighborhood of the credential if they do. The \emph{closeness} of a pair of access profiles is a sum of their weight over all credentials. In other words, two access profiles that appear frequently in small neighborhoods are the closest, those that appear frequently in large neighborhoods or less frequently in small neighborhoods are moderately close, and those that never appear together are the least close. The set of \emph{neighbors} of an access profile $u_i$ is the set of other access profiles that share a credential with $u_i$. \emph{Local homogeneity} describes how alike an access profile is to its neighbors as the average of its non-zero closeness scores. \emph{Global homogeneity} describes how homogeneous the access profiles of an anonymizing array are on average. Low global homogeneity is a reasonable objective, but keeping the maximum local homogeneity below an acceptable value is important for privacy preservation for individual subjects.

Formally, let $\mathcal{U}$ be a set of $N$ access profiles, and let $\mathcal{C}$ be the set of all credentials. Given a credential $c \in \mathcal{C}$, define the neighborhood of $c$  as $$\rho(c) = \{ u_i : u_i \text{ possesses } c, u_i \in \mathcal{U} \}.$$ Define the weight of access profiles $u_i$ and $u_j$ on credential $c$ as 
$$weight(u_i, u_j, c) =
    \begin{cases}
	    \frac{1}{|\rho(c)|} \iff \{u_i, u_j\} \subseteq \rho(c)\\
	    0 \text{ otherwise}
	\end{cases}$$
and 
$$closeness(u_i, u_j) = \sum_{c \in \mathcal{C}}weight(u_i, u_j, c)$$
The neighbors of an access profile $u_i$ are the access profiles that are in a neighborhood of $u_i$'s credentials, so define 
$$neighbors(u_i) =  \{\bigcup_{c \in \mathcal{C}}\rho(c) : u_i \in \rho(c)\}$$
define the local homogeneity metric $homogeneity(u_i)$ as 
   $$\frac{1}{|neighbors(u_i)|} \sum_{u_j \in \mathcal{U}, u_j \not = u_i} closeness(u_i, u_j)$$

and global homogeneity metric as

$$globalHomogeneity = \frac{1}{N}\sum_{u_i \in \mathcal{U}} homogeneity(u_i).$$

Considering just the sum of closeness scores without averaging over neighbors can be misleading. A high homogeneity access profile may have a few high closeness scores and many zeros, while a low homogeneity access profile may have many low closeness scores, with the result that the closeness scores of the two access profiles may be indistinguishable. By taking the average over the number of neighbors, the average closeness score of each access profile is retrieved and can be compared.

An alternative metric to closeness, $closeness'$, is the count of shared neighborhoods divided by the average neighborhood size, but it is less descriptive. Consider $closeness(u_i, u_j) = \frac{1}{2} + \frac{1}{6} = \frac{4}{6}$. The alternative metric would produce $closeness'(u_i, u_j) = \frac{2}{4} = \frac{3}{6}$. In the latter, all shared neighborhoods weigh equally, whereas in the former, the size of the neighborhood determines the weight of that neighborhood. The better metric gives a larger weight when access profiles appear together in small neighborhoods.

Division by $|neighbors(u_i)|$ requires every access profile to have at least one neighbor. This is not the same as requiring $r > 1$ which indicates that every credential has a neighborhood size of at least 2. Computation of homogeneity scores may still be useful as an interim step in an algorithm when the ideal anonymity guarantee has not yet been met. An appropriately large value for $homogeneity(u_i)$ when $|neighbors(u_i)| = 0$ needs to be determined. Choosing $\infty$ obscures $globalHomogeneity$ and prevents comparison of the array on this metric. No score with a valid number of neighbors can exceed ${k \choose t}$, so this seems like a reasonable value.

The intermediary metric, $closeness(u_i,u_j)$ is useful for finer granularity of analysis. An access profile with the highest homogeneity possible has $N-r$ access profiles $u_j$ such that $closeness(u_i,u_j)=0$ and $r-1$ access profiles $u_\ell$ with $closeness(u_i,u_\ell)=\frac{{k \choose t}}{r}$. The histogram of closeness scores for a high homogeneity access profile is bimodal. Conversely, a low homogeneity access profile has a histogram of closeness scores with a mode around $\frac{{k \choose t}}{N}$. It is likely the case that the variance of high homogeneity is higher than the variance of low homogeneity. This level of granularity may not be necessary, but retaining closeness scores might be useful when attributes can be resampled to break high homogeneity by identifying specific access profiles that are too similar. 

This level of granularity may not be required, and in this case, computation of the closeness scores can be bypassed by noting that for every credential $c$ of $u_i$ with neighborhood $|\rho(c)|$, $|\rho(c)|-1$ neighbors contribute $\frac{1}{|\rho(c)|}$ to $u_i$'s homogeneity score. This fact explains why it is important to not include $closeness(u_i,u_i)$, as doing so  reduces the homogeneity metric to $\frac{{k \choose t}}{|neighbors(u_i)|}$ for all $u_i$. This is essentially just a measure of the total number of neighbors for each $u_i$ since ${k \choose t}$ is fixed for all. A measure of the number of neighbors cannot indicate whether $u_i$ shares every neighborhood with the same set of neighbors or if it shares different neighborhoods with subsets of its neighbors. As an example, suppose ${k \choose t} = 4$ and suppose access profile $u_i$ shares all credentials with eight other access profiles. Then $homogeneity(u_i) = \frac{4(8)\frac{1}{9}}{8} = \frac{4}{9}$. Suppose another access profile $u_j$ shares each credential with two other access profiles for a total of eight neighbors. Then $homogeneity(u_j) = \frac{4(2)\frac{1}{3}}{8} = \frac{1}{3}$. As desired, $u_i$ has a higher homogeneity score as it always appears with the same profiles while $u_j$ has a slightly lower homogeneity score. If $closeness(u_i,u_i)$ is included, both scores are $\frac{4}{8}$. 

An access profile with the largest homogeneity score possible always appears with the same neighbors in the smallest neighborhoods allowed. Let $u_i$ be such an access profile appearing with the same $r-1$ neighbors for each of its credentials. Then it is the case that $weight(u_i, u_j, c) = \frac{1}{r}$ for the ${k \choose t}$ credentials of $u_i$, and $closeness(u_i, u_j) = {k \choose t}\frac{1}{r}$ for $r-1$ access profiles $u_j$. Then, as expected, \begin{equation*}homogeneity(u_i) = \frac{(r-1){k \choose t}\frac{1}{r}}{r-1} =\frac{ {k \choose t}}{r}\end{equation*} as the average closeness score of $u_i$ with each of its $r-1$ neighbors. 

Parameter values relative to other parameters affect the homogeneity. As mentioned before, the global homogeneity tends to decrease as $N$ increases. As credentials grow in size -- as $t$ approaches $k$ for fixed $N$ -- the number of neighbors expected for an access profile decreases and therefore the expected homogeneity score increases. That is, the closer a credential becomes to being a full access profile, the less other access profiles can be used to obscure the identity of the subject. If an access profile has a large number of neighbors as $t$ approaches $k$, this suggests duplication. Specifically, if an access profile has a neighbor for $t=k$, then there must be two identical access profiles. If $r > \frac{N}{v^t}$, then some credential does not appear, increasing the number of times the remaining credentials for that set of $t$ attributes appear. When $r = N$, every row is a duplicate. Therefore, the homogeneity score makes the most sense when $r$ is chosen reasonably and is useful to compare two anonymizing arrays with the same parameters. 

\subsection{Homogeneity Computation}
Algorithm~\ref{Homogeneity} computes the local homogeneity of an anonymizing array, $\mathbf{A}$.

The algorithm employs the following data structures:
\begin{itemize}
\item $v^t$ lists with average size $\frac{N}{v^t}$ for a total space of $N$ where the space can be reused for each of the ${k \choose t}$ iterations;
\item a $1 \times N$ array of double precision floating point numbers to store the homogeneity scores;
\item a $N \times N$ array of booleans to store the neighbor status.
\end{itemize}
The running time is as follows:
\begin{itemize}
\item computing the lists for each set of $t$ columns requires one scan through the $N \times t$ subarray, $\mathcal{O}(N{k \choose t})$;
\item for each of the $v^t$ lists, update the value in homogeneity, $\mathcal{O}(N)$, and update the neighbors boolean array, $\mathcal{O}(N^2)$;
\item compute the count of neighbors for each of the $N$ access profiles, $\mathcal{O}(N^2)$.
\end{itemize}
The algorithm is $\mathcal{O}(N^2 + N{k \choose t})$. It is preferred to eliminate the need to count the neighbors of $u_i$, but as explained, it is required for our metric. There does not seem to be a way to count the neighbors of an access profile without comparing it to all other access profiles. 

\begin{algorithm}
\DontPrintSemicolon
\SetKwInOut{Input}{input}\SetKwInOut{Output}{output}
\Input{ $\mathbf{A}$ an $N \times k$ array, $v$, $t$}
\Output{$Homogeneity$ a $1 \times N$ array}
\Begin
{
	\For{each of the ${k \choose t}$ subsets of $t$ columns}
	{
		Create a list $\rho(c)$ for each of the $v^t$ credentials, $c$
		
		\For{each $u_i \in \rho(c)$}
		{
			Add $\frac{|\rho(c)| - 1}{|\rho(c)|}$ to $Homogeneity[u_i]$
			
			\For{each $u_j \not = u_i \in \rho(c)$}
			{
				Set $neighbors[u_i][u_j] = true$
			}
		}
	}
	$numNeighbors[u_i]$ = count true values of $neighbors[u_i]$  
	
	Set $Homogeneity[u_i] =  \frac{Homogeneity[u_i]}{numNeighbors[u_i]}$
}
\caption{Compute Homogeneity \label{Homogeneity}}
\end{algorithm}

\subsection{Homogeneity Examples}

\begin{figure*}[h!]
    \centering
    \begin{subfigure}[b]{0.5\textwidth}
        \centering
\begin{tabular}{l | c c c}
\toprule
& $a_1$ & $a_2$ & $a_3$\\
\midrule
1 & 0 & 0 & 0\\
2 & 0 & 0 & 1\\
3 & 0 & 1 & 0\\
4 & 0 & 1 & 1\\
5 & 1 & 0 & 0\\
6 & 1 & 0 & 1\\
7 & 1 & 1 & 0\\
8 & 1 & 1 & 1\\
\bottomrule
\end{tabular}
\caption{Low Homogeneity, $\textsf{AA}(8;2,2,3,(2^3))$}
\label{fig:full8}
    \end{subfigure}%
    ~ 
    \begin{subfigure}[b]{0.5\textwidth}
        \centering
  \includegraphics[height=4.5cm]{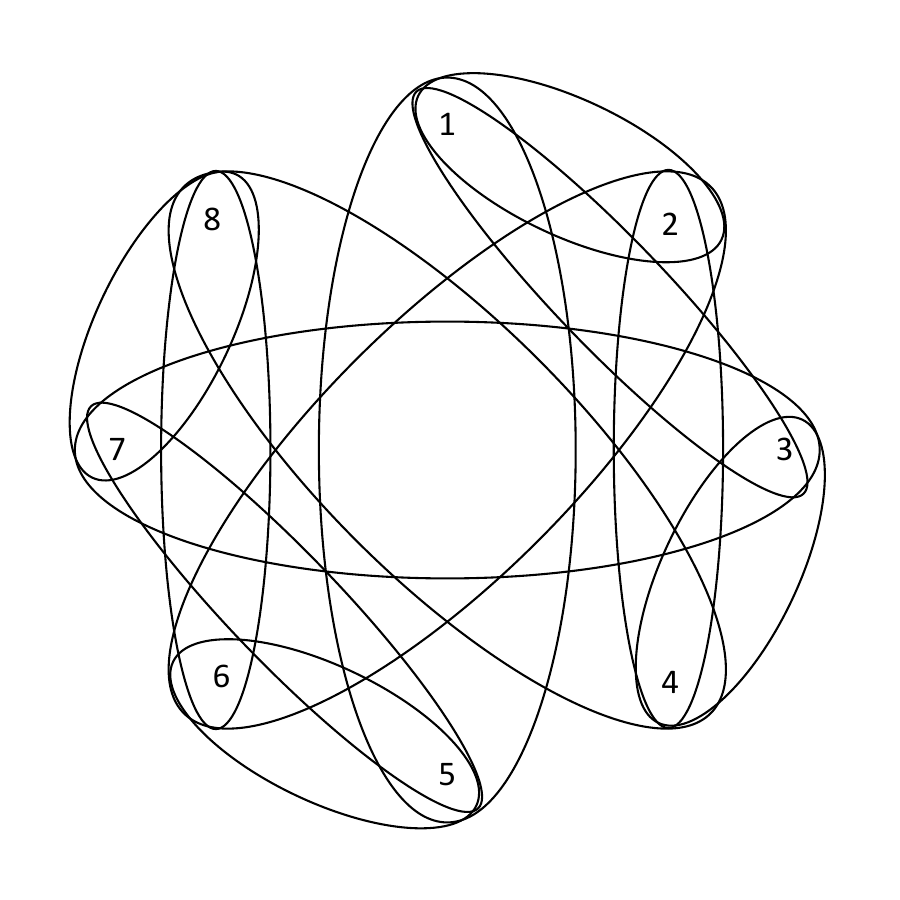}
  \caption{Low homogeneity array multi-hypergraph}
  \label{lowgraph}
    \end{subfigure}
    \caption{Low homogeneity}
\end{figure*}

\begin{figure*}[h!]
    \centering
    \begin{subfigure}[b]{0.5\textwidth}
        \centering
\begin{tabular}{l | c c c}
\toprule
 & $a_1$ & $a_2$ & $a_3$\\
\midrule
1 & 0 & 0 & 1\\
2 & 0 & 1 & 0\\
3 & 1 & 0 & 1\\
4 & 1 & 1 & 0\\
5 & 0 & 0 & 1\\
6 & 0 & 1 & 0\\
7 & 1 & 0 & 1\\
8 & 1 & 1 & 0\\
\bottomrule
\end{tabular}
\caption{Medium Homogeneity $\textsf{AA}(8;2,2,3,(2^3))$}
\label{fig:frac8}
    \end{subfigure}%
    ~ 
    \begin{subfigure}[b]{0.5\textwidth}
        \centering
    \includegraphics[height=4.5cm]{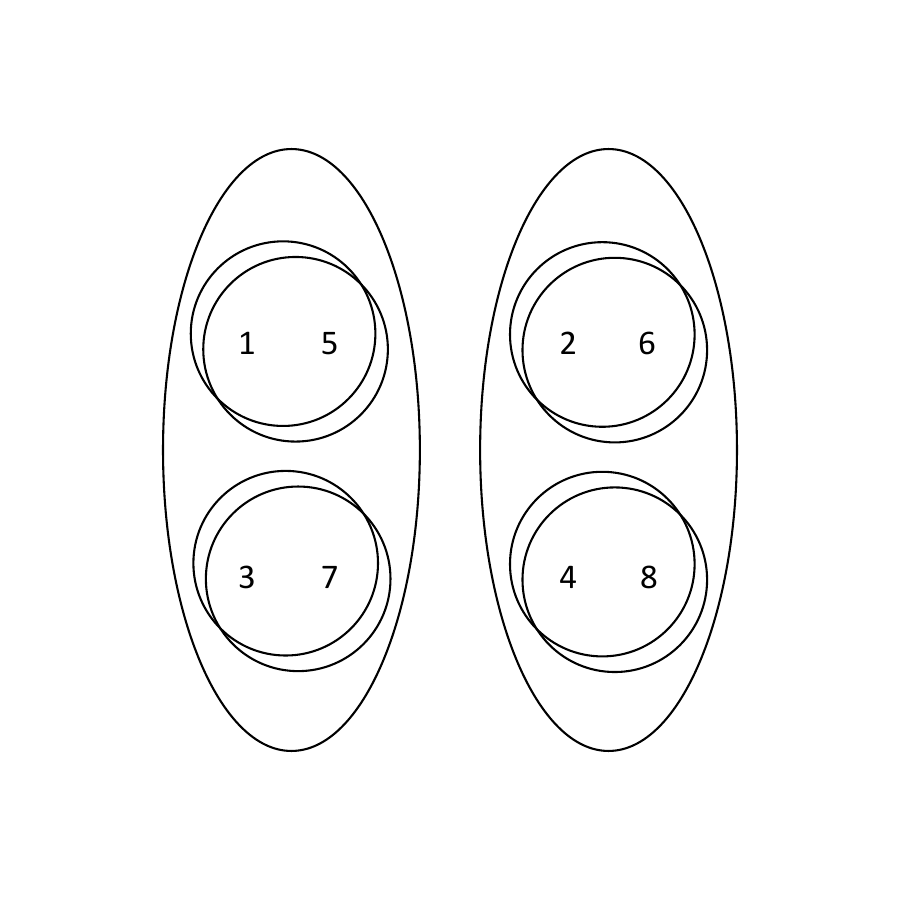}
  \caption{Medium homogeneity array multi-hypergraph}
  \label{mediumgraph}
  
    \end{subfigure}
    \caption{Medium homogeneity}
\end{figure*}

\begin{figure*}[h!]
    \centering
    \begin{subfigure}[b]{0.5\textwidth}
        \centering
\begin{tabular}{l | c c c}
\toprule
 & $a_1$ & $a_2$ & $a_3$\\
\midrule
1 & 0 & 0 & 0\\
2 & 0 & 0 & 0\\
3 & 1 & 1 & 1\\
4 & 1 & 1 & 1\\
5 & 1 & 1 & 1\\
6 & 1 & 1 & 1\\
7 & 1 & 1 & 1\\
8 & 1 & 1 & 1\\
\bottomrule
\end{tabular}
\caption{High Homogeneity, $\textsf{AA}(8;2,2,3,(2^3))$}
\label{fig:low8}
    \end{subfigure}%
    ~ 
    \begin{subfigure}[b]{0.5\textwidth}
        \centering
      \includegraphics[height=4.5cm]{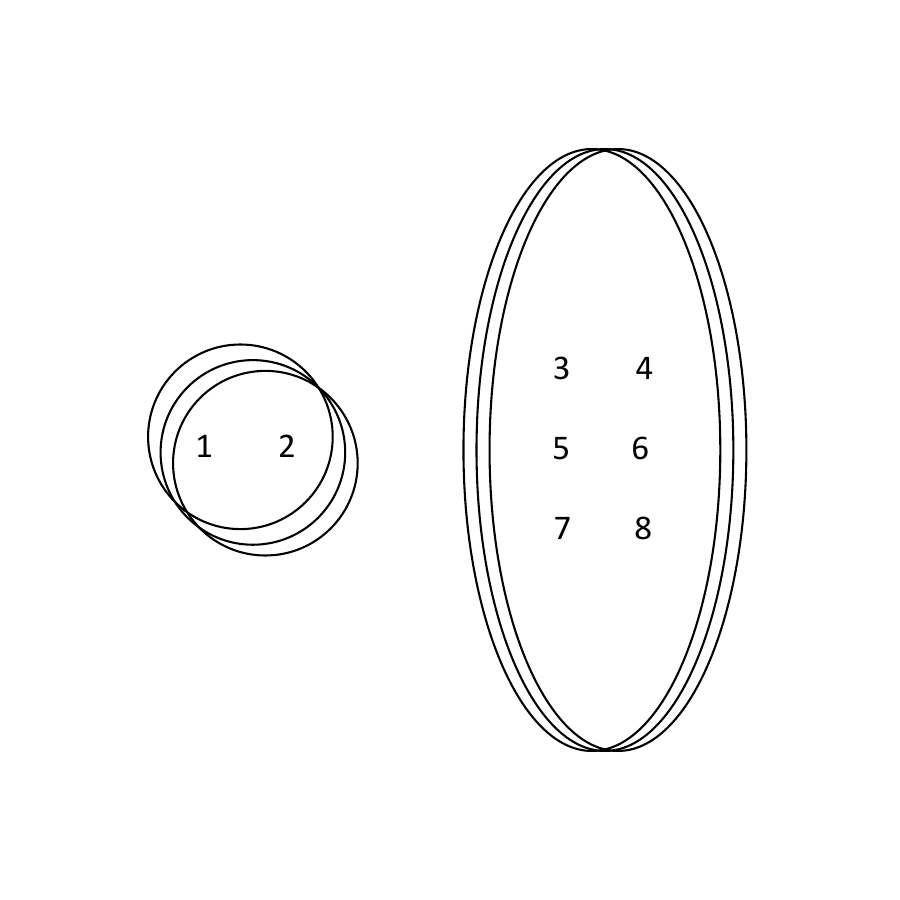}
  \caption{High homogeneity array multi-hypergraph}
    \label{highgraph}
    \end{subfigure}
    \caption{High homogeneity}
\end{figure*}

The three arrays in this section have the same parameters $N, r, t, k, v$. They do not have the same credentials appearing. The minimum, maximum, and global homogeneity scores for each array are in Table~\ref{table:scores}. A high homogeneity score is 1.5. The first, built from a $2^3$ full factorial design, is as diverse as possible with a different access profile in every row (Figure~\ref{fig:full8}). The second is  built from a $2^{3-1}$ fractional factorial with two replicates where the last four rows are copies of the first four (Figure~\ref{fig:frac8}). The third has two replicates of one access profile and six replicates of another (Figure~\ref{fig:low8}). 

The first two arrays have a uniform local homogeneity score for every access profile. The small group in the third array has a much higher homogeneity score than the large group. Even though the large group contains identical access profiles, it is not considered to be highly homogeneous, relative to the small group. While the large group can be tracked, it is anticipated that the impact of tracking to individual subjects is smaller due to the obscurity provided by the size of the group. 

Figures~\ref{lowgraph}, ~\ref{mediumgraph}, and ~\ref{highgraph} show the multi-hypergraph representations of these arrays. Access profiles are vertices labeled by the corresponding row index in the array, and an edge surrounds a set of vertices if the set of access profiles share a credential. Each vertex has degree three. The number of vertices contained in an edge varies. This representation demonstrates the clustering behavior that the homogeneity metrics are designed to measure. The edges of the hypergraph representation of the low homogeneity anonymizing array have only two vertices each and no two vertices share more than one edge. The edges of the hypergraph representation of the medium homogeneity anonymizing array are primarily concentrated in four groups of two vertices, with one edge each spanning four vertices. The edges of the hypergraph representation of the high homogeneity anonymizing array are concentrated in two groups of vertices.

\begin{table}
\caption{Homogeneity Scores}
\label{table:scores}
\begin{center}
\begin{tabular}{l l l l}
\toprule
Array & min & max & global\\
\midrule
Low Homogeneity & .5 & .5 & .5\\
Medium Homogeneity & .583 & .583 & .583\\
High Homogeneity & .5 & 1.5 & .75\\
\bottomrule
\end{tabular}
\end{center}
\end{table}


\section{Discussion}\label{section:discussion}
\paragraph{$\kappa$-Anonymity for Statistical Databases} Our work is related to the problem of
anonymizing the data in statistical databases to allow records to be released without revealing sensitive or identifying information about the individuals corresponding to the records while preventing loss of statistical value of the aggregate data. There, the privacy goal is to prevent the use of ``quasi-identifiers'' in linking records from the table being released to other databases. If linking occurs, individuals are re-identified in the released data and a ``sensitive attribute'' of one or more individuals is disclosed. While anonymizing a set of access profiles for attribute-based authorization and anonymizing records in a statistical database both share the goal of providing anonymity up to a set degree, the applications are different and so the techniques tend to be different. In the $\kappa$-anonymity problem, when the anonymized table is released, attacks attempt to determine the individual associated with a record or some sensitive information about an individual, such as the likelihood that individual A has attribute B. In our problem, all attributes of subjects are assumed to be known to the system when subjects register. Instead, we seek to prevent the system from identifying the subject presenting a credential at the moment of an access control decision. 

In the $\kappa$-anonymity literature, $k$ is the anonymity degree, while in the combinatorial array literature, $k$ is the number of columns in an array. We use $k$  for the columns of an anonymizing array and the number of attributes. Throughout this section, we use the symbol $\kappa$ for $\kappa$-anonymity to avoid confusion. 

An anonymizing array satisfies the property of ``$\kappa$-anonymity'' in \cite{sweeney2002achieving} where every $t$-set of attributes is a ``quasi-identifier'' and $r = \kappa$. A relaxation of $\kappa$-anonymity, $(\kappa,\ell)$-anonymity is defined for statistical databases when $\kappa$-anonymity causes excessive information loss and is equivalent to our definition of $(r,t)$-anonymous \cite{stokes2012computational}. The primary methods for $\kappa$- and $(\kappa, \ell)$-anonymity are \emph{generalization}, replacing an attribute value with a less specific value based on the semantics of the domain of the attribute, and \emph{suppression}, not releasing a value at all. These are applicable to the issue of anonymizing potentially sensitive data from a statistical database. For the $\kappa$-anonymity generalization methods to be useful in the attribute-based authorization domain, there must exist a ``domain generalization hierarchy'' among values of an attribute. For example, ``Ph.D. student'' and ``Masters student'' might be generalized to ``graduate student'' and possibly further generalized to ``student.'' \emph{Symbol fusing}, or mapping two symbols of a column of an array into one, is an operation used in covering arrays that is similar to generalization and might be useful when attribute values do not have a hierarchy. It is not clear when these techniques would be applicable in access control scenarios as attributes are not typically mutable. 

The $\ell$-diversity principle is a set of properties for databases satisfying $\kappa$-anonymity that requires that ``sensitive values'' within an ``equivalence class'' of a quasi-identifier, what we refer to as the neighborhood of a credential, appear not too frequently \cite{machanavajjhala2006ell}. The purpose of $\ell$-diversity is to protect against probabilistic inference attacks. This has a similar purpose as our homogeneity metrics, but homogeneity measures relationships between access profiles rather than counting the appearance of certain attribute values. It is not clear if $\ell$-diversity is a useful metric for anonymizing arrays.

When an entire row is treated as a quasi-identifier, $\kappa$-anonymity requires that $\kappa$ identical rows exist. The computational complexity of achieving $\kappa$-anonymity of a matrix with $N$ rows by suppression, replacing attribute values with $\star$, is given in \cite{bredereck2014effect}. In their work, \emph{input (output) homogeneity} is the number of different input (output) rows. In our problem, we require $r$ identical credentials, but we prefer that entire access profiles not be identical. Identical access profiles lead to high homogeneity, which is undesirable in attribute-based authorization as it may lead to tracking of a group of access profiles. This is similar to our problem when a credential is an entire row, $t=k$. It is not clear if their results generalize to our problem when $t < k$ and where the sets of attributes we seek to anonymize overlap necessarily.

\paragraph{Multiple Attribute Authorities}
Our model assumes that all attributes are known to the system and are issued by the same authority, thus any set of $t$ attributes constitutes a credential. In real world scenarios, it may be the case that different attributes are issued by different authorities \cite{chase2007multiauthority}. The implementation of a credential -- a card preloaded with attribute values, a certificate proving possession, or a private key in CP-ABE -- may dictate that only attributes from the same authority appear together in a credential. That is, their attribute sets are disjoint. Anonymizing arrays are $(r,t)$-anonymous for any $t$-set of attributes. Anonymizing arrays cover all of the potential credentials, but in a multi-authority case, this may be unnecessary. This is not the same as a hard constraint; the attributes from two authorities can and likely do appear in the same access profile, but we do not need to provide $r$ repetitions of credentials that cross boundaries of different authorities as they are never used in a policy. One solution here is to add an additional type of soft constraint, ``don't care,'' that does not require those credentials to appear exactly zero or $r$ times. 

\paragraph{Numerical Attributes}
The set system nature of anonymizing arrays is best suited when attributes are categorical. When attributes are numerical, such as $age$, and a policy can be created such as $age > 30$, it is not clear how best to extend anonymizing arrays to work in this case. One solution is to create attributes that are bracketed age ranges, such as $age=\{[20,29], [30,39], [40,49], \ldots\}$ and then list policies again in disjunctive normal form, e.g., $[30,39] \lor [40,49]$. Another solution is to enumerate all possible acceptable ages as attribute values. This works when the values are discrete, but not when they are continuous.

\paragraph{Least Privilege}
A problem for most access control systems is \emph{least privilege} or only allowing the permissions necessary to complete the job functions of the subject. There is a natural parallel with the trade-off between greater privileges and greater privacy. As a subject presents more attributes, the risk of being identified increases; as proven, there is an inverse relationship between $t$ and $r$. Therefore, the subject desires to present the fewest attributes that allow access. One approach is to require credentials composed of more attributes in order to acquire greater privileges. That is, a system guarantees a degree of anonymity $r$ for credential size $t$ to grant minimum privileges. A tiered structure may be devised so that a lower degree of anonymity $r-q$ is guaranteed when the subject presents more attributes $t+s$, but the subject also receives increased privileges. 

\paragraph{Construction Algorithms}
As is the case when constructing covering arrays, it is desirable to add as few rows as possible when appending rows to an anonymizing array. Preliminary results comparing anonymizing array construction algorithms on the number of rows is in \cite{lanus2020algorithms}, but more work developing construction algorithms tailored for anonymizing arrays is needed. If the number of rows needed to achieve $(r,t)$-anonymity is not acceptable for a required value of $r$, this method then indicates that the maximum policy size $t$ allowed should be reduced in order to meet the anonymity guarantee given the number of rows allowed. Due to the inverse relationship of $r$ and $t$, the course of action -- adding rows, reducing $r$, or reducing t -- is flexible and depends on the parameter values each instance requires or will tolerate. 

\section{Conclusions}\label{section:conclusion}
Access control decisions made on the basis of attributes afford the opportunity for anonymous authorization, but do not guarantee it when the distribution of attributes allows for the composition of policies that one or few subjects possess the credentials to satisfy. We propose anonymizing arrays as a mechanism for attribute distribution so that if credentials are restricted to $t$ or fewer attributes, subjects cannot be identified with greater than $\frac{1}{r}$ probability. We provide an algorithm that computes $r$ given an anonymizing array and maximum credential size $t$. We develop metrics, local and global homogeneity, to compare two anonymizing arrays on the same parameters, and we provide an algorithm to compute homogeneity scores. Finally, we describe open problems and directions to further develop the ideas we propose.

\bibliographystyle{model1-num-names}
\bibliography{AA}

\begin{thebibliography}{25}
\expandafter\ifx\csname natexlab\endcsname\relax\def\natexlab#1{#1}\fi
\providecommand{\url}[1]{\texttt{#1}}
\providecommand{\href}[2]{#2}
\providecommand{\path}[1]{#1}
\providecommand{\DOIprefix}{doi:}
\providecommand{\ArXivprefix}{arXiv:}
\providecommand{\URLprefix}{URL: }
\providecommand{\Pubmedprefix}{pmid:}
\providecommand{\doi}[1]{\href{http://dx.doi.org/#1}{\path{#1}}}
\providecommand{\Pubmed}[1]{\href{pmid:#1}{\path{#1}}}
\providecommand{\bibinfo}[2]{#2}
\ifx\xfnm\relax \def\xfnm[#1]{\unskip,\space#1}\fi
\bibitem[{Hu et~al.(2013)Hu, Ferraiolo, Kuhn, Friedman, Lang, Cogdell,
  Schnitzer, Sandlin, Miller, Scarfone et~al.}]{hu2013guide}
\bibinfo{author}{V.~C. Hu}, \bibinfo{author}{D.~Ferraiolo},
  \bibinfo{author}{D.~R. Kuhn}, \bibinfo{author}{A.~R. Friedman},
  \bibinfo{author}{A.~J. Lang}, \bibinfo{author}{M.~M. Cogdell},
  \bibinfo{author}{A.~Schnitzer}, \bibinfo{author}{K.~Sandlin},
  \bibinfo{author}{R.~Miller}, \bibinfo{author}{K.~Scarfone}, et~al.,
\newblock \bibinfo{title}{Guide to attribute based access control (abac)
  definition and considerations (draft)},
\newblock \bibinfo{journal}{NIST special publication} \bibinfo{volume}{800}
  (\bibinfo{year}{2013}).
\bibitem[{Kuhn et~al.(2010)Kuhn, Coyne, and Weil}]{kuhn2010adding}
\bibinfo{author}{D.~R. Kuhn}, \bibinfo{author}{E.~J. Coyne},
  \bibinfo{author}{T.~R. Weil},
\newblock \bibinfo{title}{Adding attributes to role-based access control},
\newblock \bibinfo{journal}{Computer} \bibinfo{volume}{43}
  (\bibinfo{year}{2010}) \bibinfo{pages}{79--81}.
\bibitem[{Sahai and Waters(2005)}]{sahai2005fuzzy}
\bibinfo{author}{A.~Sahai}, \bibinfo{author}{B.~Waters},
\newblock \bibinfo{title}{Fuzzy identity-based encryption},
\newblock in: \bibinfo{booktitle}{Annual International Conference on the Theory
  and Applications of Cryptographic Techniques},
  \bibinfo{organization}{Springer}, \bibinfo{year}{2005}, pp.
  \bibinfo{pages}{457--473}.
\bibitem[{Goyal et~al.(2006)Goyal, Pandey, Sahai, and
  Waters}]{goyal2006attribute}
\bibinfo{author}{V.~Goyal}, \bibinfo{author}{O.~Pandey},
  \bibinfo{author}{A.~Sahai}, \bibinfo{author}{B.~Waters},
\newblock \bibinfo{title}{Attribute-based encryption for fine-grained access
  control of encrypted data},
\newblock in: \bibinfo{booktitle}{Proceedings of the 13th ACM conference on
  Computer and communications security}, \bibinfo{organization}{Acm},
  \bibinfo{year}{2006}, pp. \bibinfo{pages}{89--98}.
\bibitem[{Bethencourt et~al.(2007)Bethencourt, Sahai, and
  Waters}]{bethencourt2007ciphertext}
\bibinfo{author}{J.~Bethencourt}, \bibinfo{author}{A.~Sahai},
  \bibinfo{author}{B.~Waters},
\newblock \bibinfo{title}{Ciphertext-policy attribute-based encryption},
\newblock in: \bibinfo{booktitle}{Security and Privacy, 2007. SP'07. IEEE
  Symposium on}, \bibinfo{organization}{IEEE}, \bibinfo{year}{2007}, pp.
  \bibinfo{pages}{321--334}.
\bibitem[{Portnoi and Shen(2016)}]{portnoi2016location}
\bibinfo{author}{M.~Portnoi}, \bibinfo{author}{C.-C. Shen},
\newblock \bibinfo{title}{Location-enhanced authenticated key exchange},
\newblock in: \bibinfo{booktitle}{2016 International Conference on Computing,
  Networking and Communications (ICNC)}, \bibinfo{organization}{IEEE},
  \bibinfo{year}{2016}, pp. \bibinfo{pages}{1--5}.
\bibitem[{Ni et~al.(2010)Ni, Bertino, Lobo, Brodie, Karat, Karat, and
  Trombeta}]{ni2010privacy}
\bibinfo{author}{Q.~Ni}, \bibinfo{author}{E.~Bertino},
  \bibinfo{author}{J.~Lobo}, \bibinfo{author}{C.~Brodie},
  \bibinfo{author}{C.-M. Karat}, \bibinfo{author}{J.~Karat},
  \bibinfo{author}{A.~Trombeta},
\newblock \bibinfo{title}{Privacy-aware role-based access control},
\newblock \bibinfo{journal}{ACM Transactions on Information and System Security
  (TISSEC)} \bibinfo{volume}{13} (\bibinfo{year}{2010}) \bibinfo{pages}{1--31}.
\bibitem[{Ardagna et~al.(2008)Ardagna, Cremonini, De~Capitani~di Vimercati, and
  Samarati}]{ardagna2008privacy}
\bibinfo{author}{C.~A. Ardagna}, \bibinfo{author}{M.~Cremonini},
  \bibinfo{author}{S.~De~Capitani~di Vimercati}, \bibinfo{author}{P.~Samarati},
\newblock \bibinfo{title}{A privacy-aware access control system},
\newblock \bibinfo{journal}{Journal of Computer Security} \bibinfo{volume}{16}
  (\bibinfo{year}{2008}) \bibinfo{pages}{369--397}.
\bibitem[{Nishide et~al.(2008)Nishide, Yoneyama, and
  Ohta}]{nishide2008attribute}
\bibinfo{author}{T.~Nishide}, \bibinfo{author}{K.~Yoneyama},
  \bibinfo{author}{K.~Ohta},
\newblock \bibinfo{title}{Attribute-based encryption with partially hidden
  encryptor-specified access structures},
\newblock in: \bibinfo{booktitle}{International conference on applied
  cryptography and network security}, \bibinfo{organization}{Springer},
  \bibinfo{year}{2008}, pp. \bibinfo{pages}{111--129}.
\bibitem[{Servos and Osborn(2017)}]{servos2017current}
\bibinfo{author}{D.~Servos}, \bibinfo{author}{S.~L. Osborn},
\newblock \bibinfo{title}{Current research and open problems in attribute-based
  access control},
\newblock \bibinfo{journal}{ACM Comput. Surv.} \bibinfo{volume}{49}
  (\bibinfo{year}{2017}) \bibinfo{pages}{65:1--65:45}.
\bibitem[{Backes et~al.(2005)Backes, Camenisch, and
  Sommer}]{backes2005anonymous}
\bibinfo{author}{M.~Backes}, \bibinfo{author}{J.~Camenisch},
  \bibinfo{author}{D.~Sommer},
\newblock \bibinfo{title}{Anonymous yet accountable access control},
\newblock in: \bibinfo{booktitle}{Proceedings of the 2005 ACM workshop on
  Privacy in the electronic society}, \bibinfo{organization}{ACM},
  \bibinfo{year}{2005}, pp. \bibinfo{pages}{40--46}.
\bibitem[{Kolter et~al.(2007)Kolter, Schillinger, and
  Pernul}]{kolter2007privacy}
\bibinfo{author}{J.~Kolter}, \bibinfo{author}{R.~Schillinger},
  \bibinfo{author}{G.~Pernul},
\newblock \bibinfo{title}{A privacy-enhanced attribute-based access control
  system},
\newblock in: \bibinfo{booktitle}{IFIP Annual Conference on Data and
  Applications Security and Privacy}, \bibinfo{organization}{Springer},
  \bibinfo{year}{2007}, pp. \bibinfo{pages}{129--143}.
\bibitem[{Jung et~al.(2014)Jung, Li, Wan, and Wan}]{jung2014control}
\bibinfo{author}{T.~Jung}, \bibinfo{author}{X.-Y. Li},
  \bibinfo{author}{Z.~Wan}, \bibinfo{author}{M.~Wan},
\newblock \bibinfo{title}{Control cloud data access privilege and anonymity
  with fully anonymous attribute-based encryption},
\newblock \bibinfo{journal}{IEEE transactions on information forensics and
  security} \bibinfo{volume}{10} (\bibinfo{year}{2014})
  \bibinfo{pages}{190--199}.
\bibitem[{Kapadia et~al.(2007)Kapadia, Tsang, and Smith}]{kapadia2007attribute}
\bibinfo{author}{A.~Kapadia}, \bibinfo{author}{P.~P. Tsang},
  \bibinfo{author}{S.~W. Smith},
\newblock \bibinfo{title}{Attribute-based publishing with hidden credentials
  and hidden policies.},
\newblock in: \bibinfo{booktitle}{NDSS}, volume~\bibinfo{volume}{7},
  \bibinfo{organization}{Citeseer}, \bibinfo{year}{2007}, pp.
  \bibinfo{pages}{179--192}.
\bibitem[{Squicciarini et~al.(2007)Squicciarini, Trombetta, Bhargav-Spantzel,
  and Bertino}]{squicciarini2007kanonymous}
\bibinfo{author}{A.~Squicciarini}, \bibinfo{author}{A.~Trombetta},
  \bibinfo{author}{A.~Bhargav-Spantzel}, \bibinfo{author}{E.~Bertino},
\newblock \bibinfo{title}{k-anonymous attribute-based access control},
\newblock in: \bibinfo{booktitle}{International Conference on Information and
  Computer Security (ICICS’07)}, \bibinfo{year}{2007}.
\bibitem[{Lanus and Colbourn(2020)}]{lanus2020algorithms}
\bibinfo{author}{E.~Lanus}, \bibinfo{author}{C.~J. Colbourn},
\newblock \bibinfo{title}{Algorithms for constructing anonymizing arrays},
\newblock in: \bibinfo{editor}{L.~G{\k{a}}sieniec},
  \bibinfo{editor}{R.~Klasing}, \bibinfo{editor}{T.~Radzik} (Eds.),
  \bibinfo{booktitle}{Combinatorial Algorithms}, \bibinfo{publisher}{Springer
  International Publishing}, \bibinfo{address}{Cham}, \bibinfo{year}{2020}, pp.
  \bibinfo{pages}{382--394}.
\bibitem[{Watts and Strogatz(1998)}]{watts1998collective}
\bibinfo{author}{D.~J. Watts}, \bibinfo{author}{S.~H. Strogatz},
\newblock \bibinfo{title}{Collective dynamics of ‘small-world’networks},
\newblock \bibinfo{journal}{Nature} \bibinfo{volume}{393}
  (\bibinfo{year}{1998}) \bibinfo{pages}{440}.
\bibitem[{Estrada and Rodriguez-Velazquez(2005)}]{estrada2005complex}
\bibinfo{author}{E.~Estrada}, \bibinfo{author}{J.~A. Rodriguez-Velazquez},
\newblock \bibinfo{title}{Complex networks as hypergraphs},
\newblock \bibinfo{journal}{arXiv preprint physics/0505137}
  (\bibinfo{year}{2005}).
\bibitem[{Jaccard(1908)}]{jaccard1908nouvelles}
\bibinfo{author}{P.~Jaccard},
\newblock \bibinfo{title}{Nouvelles recherches sur la distribution florale},
\newblock \bibinfo{journal}{Bulletin de la Société vaudoise des sciences
  naturelles.} \bibinfo{volume}{44} (\bibinfo{year}{1908})
  \bibinfo{pages}{223--270}.
\bibitem[{Nie et~al.(2015)Nie, Wu, Niu, Kuo, Leung, and
  Colbourn}]{nie2015combinatorial}
\bibinfo{author}{C.~Nie}, \bibinfo{author}{H.~Wu}, \bibinfo{author}{X.~Niu},
  \bibinfo{author}{F.-C. Kuo}, \bibinfo{author}{H.~Leung},
  \bibinfo{author}{C.~J. Colbourn},
\newblock \bibinfo{title}{Combinatorial testing, random testing, and adaptive
  random testing for detecting interaction triggered failures},
\newblock \bibinfo{journal}{Information and Software Technology}
  \bibinfo{volume}{62} (\bibinfo{year}{2015}) \bibinfo{pages}{198--213}.
\bibitem[{Sweeney(2002)}]{sweeney2002achieving}
\bibinfo{author}{L.~Sweeney},
\newblock \bibinfo{title}{Achieving k-anonymity privacy protection using
  generalization and suppression},
\newblock \bibinfo{journal}{International Journal of Uncertainty, Fuzziness and
  Knowledge-Based Systems} \bibinfo{volume}{10} (\bibinfo{year}{2002})
  \bibinfo{pages}{571--588}.
\bibitem[{Stokes(2012)}]{stokes2012computational}
\bibinfo{author}{K.~Stokes},
\newblock \bibinfo{title}{On computational anonymity},
\newblock in: \bibinfo{booktitle}{International Conference on Privacy in
  Statistical Databases}, \bibinfo{organization}{Springer},
  \bibinfo{year}{2012}, pp. \bibinfo{pages}{336--347}.
\bibitem[{Machanavajjhala et~al.(2006)Machanavajjhala, Venkitasubramaniam,
  Kifer, and Gehrke}]{machanavajjhala2006ell}
\bibinfo{author}{A.~Machanavajjhala}, \bibinfo{author}{M.~Venkitasubramaniam},
  \bibinfo{author}{D.~Kifer}, \bibinfo{author}{J.~Gehrke},
\newblock \bibinfo{title}{$\ell$ -diversity: Privacy beyond $\kappa$
  -anonymity},
\newblock in: \bibinfo{booktitle}{22nd International Conference on Data
  Engineering (ICDE'06)(ICDE)}, volume~\bibinfo{volume}{00},
  \bibinfo{year}{2006}, p.~\bibinfo{pages}{24}.
  \DOIprefix\doi{10.1109/ICDE.2006.1}.
\bibitem[{Bredereck et~al.(2014)Bredereck, Nichterlein, Niedermeier, and
  Philip}]{bredereck2014effect}
\bibinfo{author}{R.~Bredereck}, \bibinfo{author}{A.~Nichterlein},
  \bibinfo{author}{R.~Niedermeier}, \bibinfo{author}{G.~Philip},
\newblock \bibinfo{title}{The effect of homogeneity on the computational
  complexity of combinatorial data anonymization},
\newblock \bibinfo{journal}{Data Mining and Knowledge Discovery}
  \bibinfo{volume}{28} (\bibinfo{year}{2014}) \bibinfo{pages}{65--91}.
\bibitem[{Chase(2007)}]{chase2007multiauthority}
\bibinfo{author}{M.~Chase},
\newblock \bibinfo{title}{Multi-authority attribute based encryption},
\newblock in: \bibinfo{editor}{S.~P. Vadhan} (Ed.), \bibinfo{booktitle}{Theory
  of Cryptography}, \bibinfo{publisher}{Springer Berlin Heidelberg},
  \bibinfo{address}{Berlin, Heidelberg}, \bibinfo{year}{2007}, pp.
  \bibinfo{pages}{515--534}.

\end{thebibliography}


\end{document}